\documentclass[epj]{svjour}

\usepackage{graphicx}% Include figure files
\usepackage{bm}% bold math
\usepackage{psfrag} 

 \def\Gf{Greens function} \def\Gfs{Greens functions} \def\em{\ensuremath}
\def\w{\omega} \def\e{\varepsilon}
\def\k{{\bf \vec{\bf k}}}
\begin{document}

\title{Charge gaps and quasiparticle bands of the ionic Hubbard model}

\author{Torben Jabben\inst{1} \and  Norbert Grewe \inst{1} \and Frithjof
  B.~Anders \inst{2}}
\institute{
 Institut f\"ur Festk\"orperphysik,
  Technische Universit\"at Darmstadt, D-64289 Darmstadt, Germany \and
  Fachbereich 1, Universit\"at Bremen, D-28334 Bremen, Germany
}

\date{November 15, 2004}

\abstract{
  The ionic Hubbard model on a cubic lattice is investigated using
  analytical approximations, the DMFT and Wilson's renormalization
  group for the charge excitation spectrum. Near the Mott insulating
  regime, where the Hubbard repulsion starts to dominate all energies,
  the formation of correlated bands is described. The corresponding
  partial spectral weights and local densities of states show the 
  characteristic features, of a hybridized-band structure as
  appropriate for the regime at small $U$, which at
  half-filling is known as a band insulator. In particular, a narrow
  charge gap is obtained at half-filling, and the distribution of
  spectral quasi-particle weight reflects the fundamental
  hybridization mechanism of the model.
 % end of PACS codes
}

\maketitle

\section{Introduction} Investigations of the ionic Hubbard
model have led to an ongoing debate about different types of phases
with and without broken symmetries and transitions between them
\cite{Hubbard81,PozgajcicGross2003}. Reliable information has so far
been obtained mainly for the one-dimensional case, where Quantum Monte
Carlo methods \cite{WilkensMartin2001} or the density matrix
renormalization group \cite{MedenSchoenhammer2003} are applicable.
From these it became clear, that a site dependent variation of atomic
energies can induce critical behaviour and long-ranged correlations
which are absent in the homogeneous Hubbard model. This simple version
of the model in one dimension and for the particularly interesting
case of half-filling is a correlated Mott insulator for all values $U
> 0$ of the local Coulomb repulsion, with a gap for charge excitations
but none for spin excitations. Its critical bond-bond correlations
apparently freeze out at low $T$ in a parameter range $U_{1} < U <
U_{2}$, when the model is chosen ionic. For $U$ above $U_{2}$
spin-spin correlations seem to dominate, so that in a regime, where
$U$ is sufficiently larger than the local energy variation, the same
type of Mott insulating phase is approached as in the nonionic model.

The homogeneous Hubbard model in higher dimensions, namely $d = 3$,
likewise shows an interesting phase diagram, which e.g.  was
calculated in the framework of dynamical mean field
theory\cite{ZitzlerPruschkeBulla2002}. Apart from regions possibly
destabilized by phase separation, magnetic phases prevail for large
values of $U$ near half-filling, whereas upon doping the Mott
insulator in form of a correlated paramagnet is increasingly
stabilized.  Studying the paramagnetic state can serve as a good starting
point even in the regime, where the model develops magnetic order. It contains
essential information about one particle excitations and their residual interactions.
In particular, when the low lying quasi-particles form well defined bands, one may
possibly apply concepts known from the theory of weak band magnetism and
from Stoner theory and thus describe the magnetic phase in terms of exchange
splittings of heavy quasi-particle bands, at least near the phase transition line
\cite{GreweSteglich1991,Grewe1987}.
 It is therefore desirable to investigate a correlated paramagnetic state for the Hubbard model at
intermediate $U$ and small ioniticity, where formation of low lying
quasi-particle bands can be expected via an interplay of local
correlations and inhomogeneous local energies. We will thus
concentrate on one-particle excitations in a Fermi liquid phase and
study the corresponding spectral functions via appropriate many-body
techniques. Work on possible magnetic instabilities in the correlated
bands is in progress.

In the regime near the metal insulator transition, where $U$ becomes
comparable with the bandwidth as derived from nearest-neighbour
transfer, the Hubbard model near half-filling shares an important
feature with the Anderson lattice model: Bands of heavy quasiparticles
with long lifetimes form at low temperatures. Although in both cases
this process is driven by the strong local interactions and involves
complicated many-body correlations, some features of the
quasi-particle band structure seem to be linked to properties already
inherent in the one-body terms of the Hamiltonian. This applies e.g.
to the volume of the Fermi surface due to Luttingers theorem in the
Fermi liquid state, but even more detailed structure may be preserved
as can be inferred from existing work on these models
\cite{GreweSteglich1991,StichtD'AmbrumenilKuebler1986,GeorgesKotliarKrauthRozenberg1996,HeldNekrasovBluemerAnisimovVollhardt2001}.  Particularly intriguing
is the case of the spectral composition of quasi-particles at
different wavenumbers, which one could expect to be largely determined
by the structure and symmetry of the fundamental hybridization terms
of these models. We will show that indeed this also occurs in the
ionic Hubbard model.

The motivation to link a study of the ionic Hubbard model to results for the
Anderson lattice is twofold: As will be discussed in the following, ionicity
of the Hubbard model causes  similarities concerning the Brillouin zone 
and the band structure. Furthermore, new features to be expected due to these
similarities require  a certain standard of many-body techniques,
a lecture which was originally learned in connection with the Anderson lattice.
Application of the Friedel sum rule and of the Luttinger theorem
\cite{Luttinger1960} to the Anderson lattice model sheds some light on
 the density of states (DOS) in the region of low lying quasi-particles 
 \cite{MarinAllen1979}. In particular for one
conduction band and two electrons per site the noninteracting version
of the model has a hybridization gap at the Fermi energy, and the
Fermi surface coincides with the boundary of the first Brillouin zone,
a situation which should be preserved when interactions are included.
More generally, one may argue that a coherent action of the Kondo
effect should lead to a gap at the energetic position where the many
body resonances form on each lattice site. Due to Friedels sum rule it will depend on the
localized charge whether this position is near the Fermi energy and
thus leads to a charge excitation gap. Indeed, many-body calculations
using various techniques have demonstrated that a picture of
hybridized quasi- particle bands is an appropriate description of the
low temperature regime of this model
\cite{Grewe1987,Grewe1984,FuldeKellerZwicknagl1988,Kuramoto1985,MetznerVollhardt1989}.

The Hubbard model likewise contains interactions and hybridization
terms, the latter usually being addressed as transfers to nearest
neighbours. The noninteracting version gives one simple tight binding
conduction band, and the generic case puts the Fermi level somewhere
near the band center. Correspondingly, the Fermi surface lies well
inside the first Brillouin zone and a gap is not expected in the
interacting case, too. The situation can change drastically when an
ionic Hubbard model is considered. For a bipartite, simple cubic
lattice with local one particle energies differing between the two
sublattices, the first Brillouin zone is halfed, and with one electron
per site the Fermi surface coincides with the new, smaller zone
boundary. Then it can again be expected that the symmetry-breaking
effective field produces enough Bragg scattering to lift the
degeneracy at the zone boundary and thus leads to a gap like in the
Anderson lattice with two electrons per site. We would therefore
expect on the basis of this similarity that at least near half-filling 
quasi-particle bands form at low $T$, which reveal the underlying 
hybridization mechanism. In addition, it would be interesting to find
out, whether the wave vector dependence of the corresponding matrix 
elements is preserved in the structure of quasi-particles.

This expectation, in fact, is supported by a theoretical approach,
which has proven extremely useful for such models with dominant local
interactions. The approach rests upon the picture of effective sites
\cite{Grewe1987}, which react to a surrounding medium which in turn is
formed by these same local objects. Links between them are established
via the same elementary transfer processes which lead to band
formation via tight binding of noninteracting electrons.  Scattering
of quasi-particles by the effective sites thus reflects in lowest
order just the $\k$-dependence of these hybridization matrix elements,
whereas the important part of the selfenergy remains local, i.e.
$\k$-independent. Quite a long line of approximations have been
formulated along these lines, from the ATA \cite{Grewe1984} and the
Renormalized Band Theory \cite{FuldeKellerZwicknagl1988} to the XNCA
\cite{Kuramoto1985} and the Dynamical Mean field Theory (DMFT)
\cite{MetznerVollhardt1989}, which has become very popular. It
possesses a sound justification and has formed the basis of extensive
studies of Hubbard and Anderson models
\cite{GeorgesKotliarKrauthRozenberg1996}. Generally, quasi-particle
bands of the type outlined above are derived with these methods
supporting our view and furnishing a good prospect for generalizations
e.g. to the ionic Hubbard model.

Early approaches to
the Anderson lattice model like the ATA and the LNCA used simplified
versions of the Non-Crossing Approximation (NCA) at finite $U$ for the
calculation of scattering processes by the effective sites. They
clearly pointed to a quasi-particle band structure with a
hybridization gap. A more consistent description of the effective
medium, as contained in the XNCA, put this in doubt for a while,
because subtle cancellations between local and nonlocal contributions
to the selfenergy do not occur in the right way due to shortcomings of
the NCA. When combined with a better local method the selfconsistency
cycle of the XNCA produces the gap, too. In combination e.g. with the
Numerical Renormaliztion Group (NRG) \cite{PruschkeBullaJarrel2002}
this has become a reliable tool, at least for low temperatures and
excitation energies. The simplified NCA, as an analytical tool, 
can be expected to become still more useful here, when an extended or  full
version is manageable in numerical calculations. In connection with
 the Hubbard model this conceptual frame is known as the
Dynamical Mean Field Theory (DMFT) and will be used for the essential 
calculations in section 3 of this paper. Since less ambitious approximations
in the spirit of Hubbard I or with analytical impurity solvers like SNCA in the DMFT-cycle are
helpful to understand the initial stages of correlation effects and to span
Luttinger's szenario from the noninteracting to the fully interacting case, we
first give some elementary calculations demonstrating rough overall features
of the expected hybridized band structure. The reader mainly interested in
the correct final form of the results should proceed to the second half of
section 3.

In the following section 2 we will introduce our model Hamiltonian
and discuss shortly the techniques used to solve the model. The
transition from the energy levels of isolated ions towards the fully
developed picture of correlated quasi-particle bands will be
performed in two steps: At first, it is instructive to study the
splitted bands furnished by a generalized Hubbard I-approximation,
called Free Theory in the following for brevity \cite{This term},
 which neglects all two particle correlations except for the purely local ones.
It allows for a discussion of partial densities of states (DOS),
interaction - and hybridization - induced band splittings and van Hove
singularities in the absence of lifetime effects. Section 3 proceeds to a 
generalized DMFT-scheme, which includes  the complete set of local
correlations and is self-consistent on the lattice.  Calculations are carried out with an
analytical impurity solver, a simplified version of the finite $U$ -
NCA \cite{PruschkeGrewe1989}, and alternatively with the NRG. Whereas
the former allows for a qualitative study of lifetime effects and is
particularly useful at higher excitation energies, a clear cut picture
of hybridization effects in the region of the many-body resonance is
only achieved with the latter. The final section 4 contains a short
discussion and concluding remarks about future perspectives.

\section{The bipartite Hubbard model and correlated bands}
We build our model with two sorts of $s$-shells with one particle
energies $\varepsilon_A, \varepsilon_B = \varepsilon_A + \Delta$ and
local Colomb matrix elements $U_A, U_B$. Each of them is placed onto
one sublattice $L_A, L_B$ of a three-dimensional simple cubic lattice,
so that the nearest neighbours of one sort belong to the other. The
Hamiltonian is:
\begin{eqnarray}
  H & = & H_A + H_B + H_{AB} \quad \mbox{with} \nonumber \\
 H_\nu & = & \sum_{j \varepsilon L_\nu} ( \sum_{\sigma} \; 
\varepsilon_{\nu j \sigma} + U_\nu n_{\nu j \uparrow}
n_{\nu j \downarrow})  \nonumber \\
&  = &H^{(0)}_\nu + U_{\nu} \sum_{j} \; n_{\nu j \uparrow} 
n_{\nu j \downarrow}\; \;  (\nu = A, B),  \nonumber \\
H_{AB} & = & t \; \sum_{j \varepsilon L_A} \; \; \sum_{\ell \varepsilon L_B \, n.N. of \; j} 
a^+_{j \sigma} b_{\ell \sigma} + h. c. \; .
\end{eqnarray}

\begin{figure}[htbp]
  \centering \includegraphics[scale=0.5]{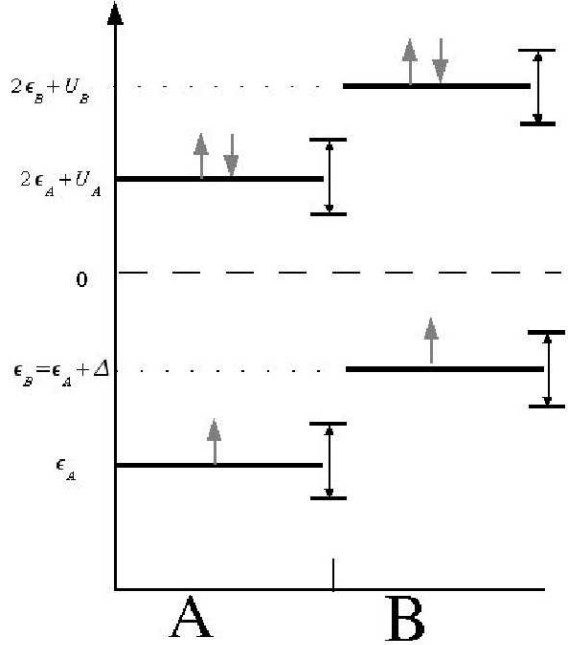}
    \caption{Scheme of the local shells and the energy regime used for the
      bipartite Hubbard model. The expected bandwidth is indicated by
      the vertical delimeters. The Fermi level lies well separated
      between the lower and upper bands.}
 \label{fig:levelscheme}
\end{figure}

We have introduced the local occupation operators $n_{Aj \sigma} =
a^+_{j \sigma} a_{j \sigma}$ and $ n_{B \ell \sigma} = b^+_{\ell
  \sigma} \, b_{\ell \sigma}$ for sites on each sublattice and
absorbed the (common) chemical potential into the one particle
energies.  A scheme of the resulting energies of local shells is shown
in Fig.~\ref{fig:levelscheme}, which also gives an impression of the
energy regime we have in mind when the indicated position of the Fermi
level and the zeroth order bandwidths are recognized.

\begin{figure} [htbp] 
  \centering \psfrag{pi/2}{$ \frac{\pi}{2a}$}
  \psfrag{-pi/2}{$-\frac{\pi}{2a}$} 
  \psfrag{e-2t}{$ \epsilon_A+2t$}
  \psfrag{e}{$\epsilon_A$} \psfrag{e+d}{$\epsilon_B$}
  \psfrag{e+d+2t}{$\epsilon_B-2t$} \psfrag{energie}{$\omega$}
  \psfrag{rho_A}{$\rho_A(k,\omega)$}
  \psfrag{rho_B}{$\rho_B(k,\omega)$} \psfrag{k}{$ k$}
  
  \includegraphics [scale=0.45]{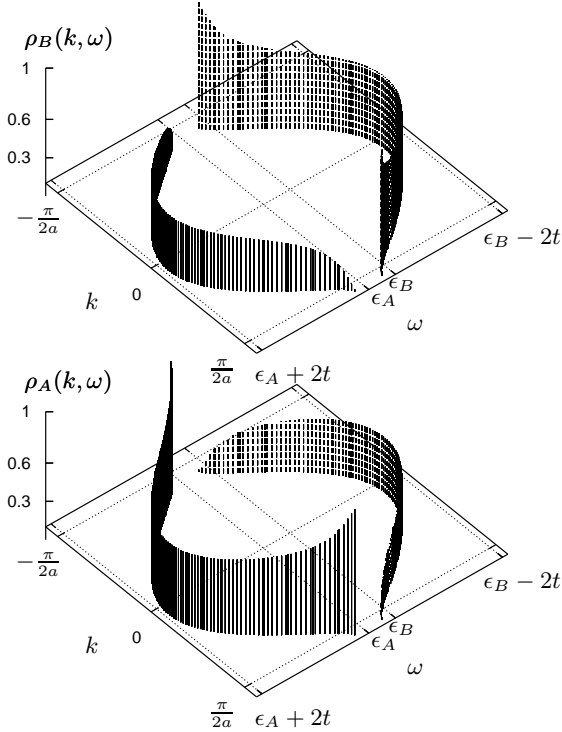}
\caption{Tight-binding approximation of the one dimensional partial
  DOS for the A and B sublattice. }
\label{fig:parttight}
\end{figure}

This atomic limit is described by the Hamiltonian $H_A + H_B$. If the
transfer term is added instead of the interactions, i.e.  when
\begin{equation}
H^{(0)} = H^{(0)}_A + H^{(0)}_B + H_{AB}
\end{equation}
is considered, a picture of hybridized tight binding bands for
noninteracting electrons emerges, which is easily visualized: Assuming
first $\Delta = 0$, the single tight binding band of a homogeneous
$(\varepsilon_A = \varepsilon_B)$ lattice is folded back corresponding
to a halfing of the Brillouin zone. The effect of $\Delta$ then shifts
the remaining original and backfolded pieces apart, producting gaps
due to Bragg-scattering between the boundaries of the new reduced
zone. Fig.~\ref{fig:parttight} gives an impression of this band
structure, using a single cosine for the dispersion function
$\varepsilon^{(0)}_{\k} = 2 t \cos(ka) \; \; (a=\mbox{lattice
  constant}, t \leq 0)$ as appropriate for one dimension.

Shown are the spikes corresponding to the partial DOS
\begin{equation}
\rho_A (\k, \omega) = 
-\frac{1}{\pi} \, Im \; G_{a_{\k \sigma},  a^+_{\k \sigma}} \; \;
(\omega + i \delta) \quad (\rho_B \; \; \mbox{analogous})
\end{equation}
derived separately for excitations on each sublattice.

It is easy to see (e.g. perturbatively with respect to $t$ or with the
equation of motion method) that the Greens functions appearing here
are given by
\begin{eqnarray}
&& G_{a_{\k \sigma},  a^+_{\k \sigma}} (z) 
 = \Bigl[G^{(0)}_A (z)^{-1} \, - \, (\varepsilon_{\k})^2 G^{(0)}_B (z) \Bigr]^{-1}  \nonumber \\
&& (G_{b_{\k \sigma},   b^+_{\k \sigma}} \mbox{analogous}),
\end{eqnarray}
where the counterparts for the noninteracting atomic limit have to be
inserted, i.e.
\begin{equation}
G^{(0)}_A (z) = [ z - \varepsilon_A]^{-1} \quad (G_B^{(0)} \; \; analogous).
\end{equation}

Whereas the added spectral weight $\rho = \rho_A + \rho_B$ would
simply exhibit two $\delta$-spikes with weight $\frac{1}{2}$ at the
energies of the upper and lower band for each fixed value of crystal
momentum $\k$ and spin $\sigma$, the distribution of the partial
spectral weights is more interesting, following
\begin{eqnarray}
&& \rho_A (\k, \omega) = \zeta^{(-)}_A (\k) \cdot \delta (\omega -
\varepsilon^{(+)}_{\k}) + \zeta^{(+)}_A (\k) \cdot \delta
(\omega - \varepsilon^{(-)}_{\k}) , \nonumber \\ 
&& \mbox{with}\quad
\zeta^{(\pm)}_A = \frac{1}{2} \pm \frac{\Delta}{2 \sqrt{\Delta^2 
+ (2 \varepsilon_{\k}^{(0)})^2}} , \nonumber \\
&&\varepsilon^{(\pm)}_{\k} = \frac{1}{2} \Bigr(\varepsilon_A + \varepsilon_B \pm
\sqrt{\Delta^2 + (2 \varepsilon_{\k}^{(0)} )^2}\; \Bigr).
\end{eqnarray}
For $\rho_B$ one has to exchange $\zeta_B^{(+)} = \zeta^{(-)}_A,
\zeta^{(-)}_B = \zeta^{(+)}_A$.  The gap at the zone boundary equals
that of the local states, i.g. $\varepsilon^{(+)} - \varepsilon^{(-)}
= \sqrt{\Delta^2 + (2 \varepsilon^{(0)}_{\k})^2} = \Delta \;
\;\mbox{for} \; \; k = \pm \frac{\pi}{2a}$, and at these two points
the bands terminate in the local energies $\varepsilon^{(+)}_{\k} =
\varepsilon_B \; \; \mbox{and}\; \; \varepsilon^{(-)}_{\k} =
\varepsilon_A \;\; \mbox{for}\; \; k = \pm \frac{\pi}{2a}$, inducing
complete repulsion of admixture from the corresponding other state.
Therefore the upper band has full $B$-weight at the zone boundaries
and less in the middle, whereas the $A$ weight goes to zero at the
boundaries and grows towards the middle of the Brillouin zone and the
other way round for the lower band. This simply reflects the shape of
the dispersion function $\varepsilon^{(0)}_{\k}$, which is zero at the
boundaries and has maximal absolute value $|2t|$ in the zone center,
since this function also gives the effective hybridization of
electrons with momentum $\k$ in transfer processes between a site and
all of its nearest neighbours.

\begin{figure} [ht] 
  \centering \psfrag{pi/2}{$ \frac{\pi}{2a}$} \psfrag{-pi/2}{$
    -\frac{\pi}{2a}$} \psfrag{pi/4}{$ \frac{\pi}{4a}$}
  \psfrag{-pi/4}{$ -\frac{\pi}{4a}$} \psfrag{z}{$ \omega$}
  \psfrag{k}{$ k$} \psfrag{rho_A}{$\rho_A(k,\omega)$}
  \psfrag{rho_B}{$\rho_B(k,\omega)$} \psfrag{energie}{$\omega$}
  
  \includegraphics [scale=0.45]{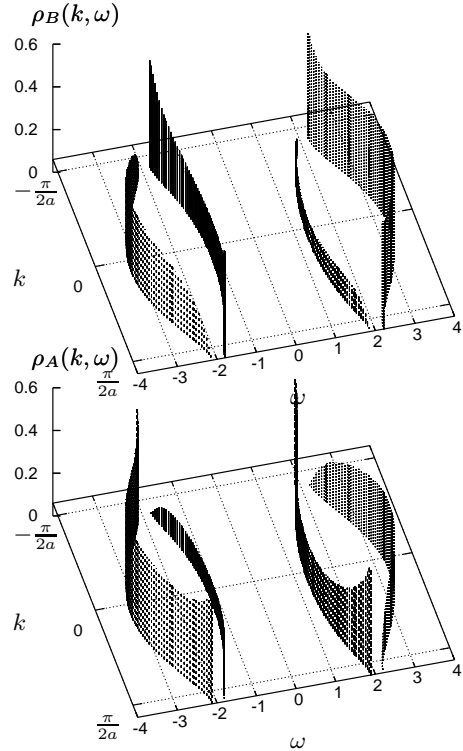}

 \caption{Partial one dimensional DOS of the A and the B sublattice
   from Free Theory calculations for 
   $\epsilon_A=-2,\epsilon_B=-1.66$, $U_A=U_B=4$. Due to the
   energy difference $\Delta=\epsilon_B-\epsilon_A$ each Hubbard band
   splits in two subbands with hybridization gaps of width $\Delta$. }
  \label{fig:partfree}
\end{figure}

It is worthwhile to collect these simple facts, since the effects are
relevant also for the more
refined band structure derived with better approximations.  A 
generalization of the  Hubbard I-approximation, named Free Theory,
can be obtained by simply substituting the Greens functions of the
isolated interacting local shells, i.e.
\begin{equation}
G^{(1)}_A (z) = \frac{\zeta^{(1)}_A}{z - \varepsilon_A} +\frac{\zeta^{(2)}_A}{z - \varepsilon_A - U_A}
\quad (G^{(1)}_B \;  \; analogous),
\end{equation}
for the $G^{(0)}_{A, B}$ given by Eq.\hspace{1mm}(5) in the explicit
expressions (4). The resulting bands are derived from a polynomial of
fourth order, the zeroes of which are easily determined numerically,
as well as the corresponding partial spectral weights, too. Here we
give the explicit dispersion for the special case $U_A = U_B \equiv U
\; \; \mbox{and} \; \; T = 0 \; \; \mbox{where} \; \; \zeta^{(1)}_A =
\zeta^{(2)}_A = \zeta^{(1)}_B = \zeta^{(2)}_B = \frac{1}{2}$ in the
situation depicted in Fig.~\ref{fig:levelscheme}:
\begin{eqnarray}
&& \varepsilon^{(m)}_{\k} = \frac{1}{2} \Bigl(  \varepsilon_A +  \varepsilon_B
 + U \pm  \nonumber \\
&&  \sqrt{\Delta^2 + U^2 + (2 \varepsilon^{(0)}_{\k})^2 \pm 2 \; \; 
\sqrt{(\Delta U)^2 + (U \varepsilon^{(0)}_{\k})^2 + 
( \varepsilon^{(0)}_{\k})^4}} \Bigl) , \nonumber \\
&& \mbox{with} \nonumber \\
&& m \equiv 1 \hat{=} (+, +), m \equiv 2 \hat{=} (+, -), \nonumber \\
&& m \equiv 3   \hat{=} (-, -), m \equiv 4 \hat{=} (-, +). 
\end{eqnarray}
The band structure is shown in Fig.~\ref{fig:partfree}, where the
spectral weights are again resolved in contribution from sublattices
$A$ and $B$. Throughout the paper, all energies will be given in units
of the hopping parameter $|t|$.

One recognizes the same distribution of weight as in the
noninteracting case before. The important new feature occuring in the
Free Theory is the interaction-induced additional splitting of the
hybridized bandstructure, formerly seen in Fig.~\ref{fig:parttight},
which is in complete agreement with the phenomenon known from Hubbards
own study: Each of the bands splits in two subbands of approximately
half the spectral weight, and differing in energy by roughly the
Coulomb repulsion. Interesting is also the corresponding local DOS,
which is obtained via summation over all $\k$. A glance at the
upcoming Fig.~\ref{fig:ABfreedmft} shows typical van Hove
singularities at the bandedges. In the present bipartite case they may
always be attributed to one of the sublattices only, since for the
other sublattices the spectral weight vanishes at the corresponding
energies. Concluding this section, we comment shortly on the validity
of Eq.\hspace{1mm}(4) and  similar generalizations of the
 Hubbard I-approximation. The
essential point is the neglect of all nonlocal correlations and
thus the restriction to the most important local interaction effect.
This is contained in the fractional form of the Greens function (7),
which is easily generalized to a more complicated local level
structure. Using this Greens function, the approximation simply
proceeds as in the noninteracting case, involves the Wick theorem and
all benefits thereof.  This may also be obtained from a decoupling in
equations of motion.

\section{Selfconistent approximations for effective sites
  and low lying quasi-particle states}

Corrections to the Free Theory
can be organized perturbatively with help of local cumulant
interactions serving as vertices. In order to tackle the infrared
problem connected with degeneracies in the metallic regime
resummations to infinite order are necessary \cite{KeiterKimball1971,Metzner1991}. 
The well established strategy of DMFT uses the concept of
effective sites, selfconsistently coupled to an environment, which is
built up by transfering electrons between these same objects.  This is
accomplished by using an appropriately highly developped technique for
an impurity problem and by feeding it with effective propagators for
electrons leaving and entering and being scattered in between by other
effective sites of the lattice. It is well known how this concept is
formally implemented in an elegant manner as a small set of equations 
\cite{Kuramoto1985,MetznerVollhardt1989}; we will just state here
its generalisation to the present situation with a few comments
concerning the role of the quantities appearing.

The two types of lattice sites experience different forms of local
one-particle excitation processes, which as before are described via a
Greens function in the form
\begin{equation}
G_A (z) \equiv G_{a_{j \sigma}, a^+_{j \sigma}} (z) = \frac{1}{2} \sum_{\k}
G_{a_{\k \sigma}, a^+_{\k \sigma}} (z) \; \; (G_B \;  analogous).
\end{equation}
For a description of the propagation process inherent in the
$\k$-dependent Greens function under the sum a division into free
transfer processes and effective local parts is helpful. This enables
one to apply the same reasoning as was used to derive
Eq.\hspace{1mm}(4) and thus gives an analogous result here:
\begin{eqnarray}
&& G_{a_{\k \sigma}, a^+_{\k \sigma}} (z) =
\bigl[ \tilde{G}_A (z)^{-1} - (\varepsilon_{\k})^2 \tilde{G}_B (z) \Bigr]^{-1} \nonumber \\
&& (G_{b_{\k \sigma}, a^+_{\k \sigma}} \; \; analogous).
\end{eqnarray}
The effective local \Gfs $\; \tilde{G}_A \; \mbox{and}\; \tilde{G}_B$
contain corrections for processes, where particles leave and enter the
site along an irreducible loop of transfers. Since such loops on the
one hand contribute to the pseudolocal Greens functions $ G_A \;
\mbox{and} \; G_B $ and on the other are produced by the free random
walk through the lattice leading to the result Eq.\hspace{1mm}(10),
these corrections are necessary to avoid overcounting.  They are
collected in quantities $ \tilde{T}_A (z) \; \mbox{and} \; \tilde{T}_B
(z) $, which enter in the following way:
\begin{equation}
\tilde{G}_A (z)^{-1} = G_A (z)^{-1} + \tilde{T}_A (z) \quad (\tilde{G}_B (z) \; \; analogous),
\end{equation}
where $G_A$ results from Eq.\hspace{1mm}(9). In this way a closed
selfconsistency cycle is obtained, in the present form with coupled
$A$- and $B$-sublattices, which has to be supplemented as mentioned by
an impurity theory for $G_A$ and $G_B$.  In the framework of the DMFT
\cite{MetznerVollhardt1989,PruschkeJarrellFreericks1995,GeorgesKotliarKrauthRozenberg1996},
the quantities $\tilde{T}_{A/B} (z) $ are usually called the dynamical
Weiss fields of the theory. Recently it was pointed out
\cite{BrandtMielsch89} that they determine how much the effective local
inverse \Gf $\; \tilde{G}_{A/B} (z)^{-1}$ has to be deformed to
reproduce the $\k$-summed lattice \Gf $\; G_{A/B}$ .

Analytical impurity solvers have been derived from the noncrossing
approximation (NCA) of the Anderson impurity model \cite{Grewe1983}.
Their virtues and shortcomings are well known
\cite{KuramotoKojima1984}. A full version of the NCA for finite
Coulomb repulsion has been presented long ago
\cite{PruschkeGrewe1989}, but for a quick orientation or for
complicated applications usually a simplified version, the SNCA is
employed, which needs much less numerical effort. Also improvements of
the full NCA have been presented \cite{AndersGrewe1994,KrohaWoelfleCosti1997}, 
which however consume even more time and
effort. The SNCA, and to a lesser degree the full NCA, loose their
relliability in the regime of very low temperatures and excitation
energies. They describe, however, rather well the overall structure
and can be used down to temperatures and excitation energies of the
order of the many-body scale, connected with the infrared problem, and
somewhat below. In particular the full NCA reproduces this
nonperturbative scale with quantitative accuracy
\cite{PruschkeGrewe1989}. We have employed here the SNCA in order to
obtain a first test of the selfconsistency cycle outlined in
Eqs.\hspace{1mm}(9) to (11).  Thereby, $G_A$ and $G_B$ are calculated
using the loops $\tilde{T}_A$ and $\tilde{T}_B$ as input, afterwards
one obtains $\tilde{G}_A$ and $\tilde{G}_B$ using (11) with these $G_A
$ and $\tilde{T}_A$, likewise for $\tilde{G}_B$ , and finally a new
$G_A$ and $G_B$ is produced with (10) inserted into (9). These $G_A$
and $G_B$ give rise to new loop-expressions $\tilde{T}_A$ and
$\tilde{T}_B$, again via Eq.\hspace{1mm}(11), which in turn allow for
a new impurity calculation of $G_A$ and $G_B$.

\begin{figure} [ht] 
  \centering \psfrag{rho}{$ \rho_{AB}(\omega)$}
  \psfrag{energie}{$\omega$} \psfrag{E_A=-2.07}{$\epsilon_A=-2.07$}
  \psfrag{E_B=-1.89}{$\epsilon_B=-1.89$} \psfrag{U=3.96}{$U_A=3.96$}
  \psfrag{d=0.18}{$U_B=3.98$} \psfrag{1/T=20}{$\beta\rightarrow
    \infty$}
 
  \psfrag{rho_A}{$\rho_A(\omega)$} \psfrag{rho_B}{$\rho_B(\omega)$}
  \psfrag{rho_AB}{$\rho_{AB}(\omega)$}
  
  \includegraphics
  [scale=0.45]{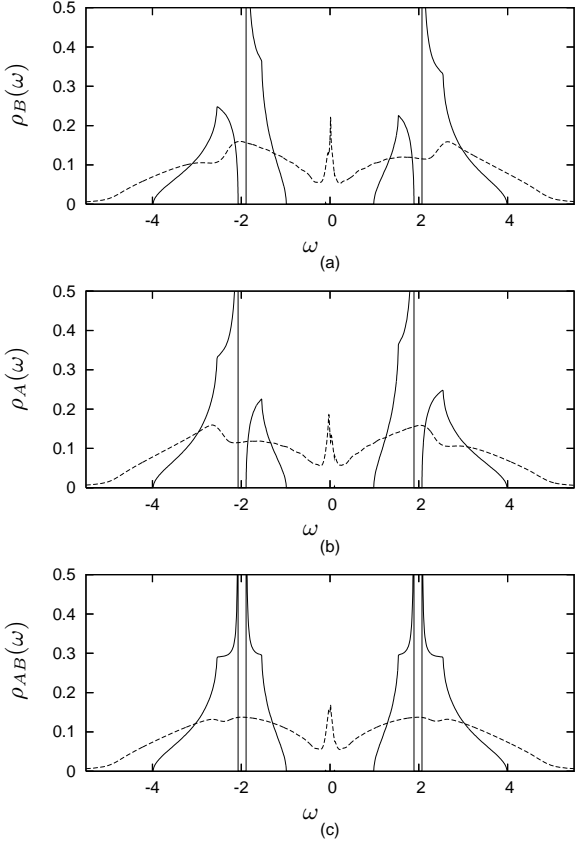}

 \caption{Comparison of the three dimensional local DOS obtained from
   Free Theory (solid line) and DMFT+SNCA (dashed line) for
   $\epsilon_A=-2.07$, $\epsilon_B=-1.89$,
   $U_A=U_B=3.96$, $\beta=20$.}
 \label{fig:ABfreedmft}

\end{figure}

In Fig.~\ref{fig:ABfreedmft} results of this iterative procedure are
shown and compared to the corresponding quantity calculated in the
Free Theory Fig.~\ref{fig:ABfreedmft}(c) contains the complete local
DOS;
\begin{eqnarray}
&& \rho_{AB} (\omega) = - \frac{1}{2\pi} Im [G_A (\omega + i \delta) +G_B (\omega + i \delta)]
 \equiv \nonumber \\
&& \equiv  \frac{1}{2} [\rho_A (\omega) + \rho_B (\omega)].
\end{eqnarray}
and compares it to the corresponding quantity calculated in the Free
Theory.

It is surprising, how much lifetime effects, i.e. the broadening of
DOS peaks at fixed $\k$, smear out the prominent features of the DOS,
which have been discussed in Section 2. The hybridization gaps now
only show up as shallow dips and the van Hove singularities have
disappeared. The prospect, however, to obtain hybridization-induced
gaps in the narrow quasi-particle peak around the Fermi level $\omega
= 0$ are unbroken: The Fermi liquid state guarantees divergent
lifetimes for $T, \omega \rightarrow 0$. It seems thus encouraging,
that the calculation indeed shows some precursor of splitting near
$\omega = 0$, which is more pronounced in the partial DOS shown in
Figs.~\ref{fig:ABfreedmft}(a) and ~\ref{fig:ABfreedmft}(b).
Unfortunately, our selfconcistency cycle with the simplified SNCA
tends to become unstable in the most interesting region and does not
allow for a more precise investigation of this interesting effect. One
should notice from an inspection of Fig.~\ref{fig:ABfreedmft},
however, that some of the characteristic features of the bipartite
lattice, which have been discussed before, survive even in such a
locally complete calculation, although only in a smoothed form.

In order to answer reliably the question for hybridization structure
in the low energy quasi-particle domain we have finally employed
Wilsons renormalization group as the impurity solver in the
selfconsistency cycle
\cite{Wilson75,KrishWilWilson80b,BullaHewsonPruschke98} defined above.
In the meantime, this has become a standard procedure, which
essentially improves on the low energy-low temperature results, but
due to its logarithmic partition of excitation energies around $\omega
= 0$ tends to suppress important DOS structure elsewhere
\cite{BullaPruschkeHewson97}. This latter statement is substantiated
in Fig.\hspace{1mm}5, viewing the smooth and featureless main
resonances near $\varepsilon_A$ and $\varepsilon_A + U_A$, likewise
for $B$. At the Fermi level $\omega = 0$, on the other hand, this $T =
0$ calculation reveals a clear and complete hybridization gap in the
quasi-particle DOS at the lowest excitation energies. Although the
present calculation uses parameter values $U_A = U_B = 3.96,
\varepsilon_A = -2.07, \varepsilon_B = -1.89 $ near the Mott
transition, we expect at least for half-filling this gap to exist at
the Fermi level for all values of $U$ in the Fermi-liquid state, as we
have motivated above with reference to the Luttinger theorem.
\newline\indent
\begin{figure} [ht] 
  \centering
  
  \psfrag{omega}{$ \omega$} \psfrag{rhoAB}{$\rho_{AB}(\omega)$}
  \psfrag{rhoA}{$\rho_{A}(\omega)$} \psfrag{rhoB}{$\rho_{B}(\omega)$}
  \psfrag{E_A=-1.5}{$\epsilon_A=-1.5$}
  \psfrag{E_B=-1.3}{$\epsilon_B=-1.3$} \psfrag{UA=3}{$U_A=3$}
  \psfrag{UB=3}{$U_B=3$} \psfrag{T=0}{$\beta\rightarrow \infty$}
 
  \psfrag{energie}{$\omega$} \includegraphics
  [scale=0.35]{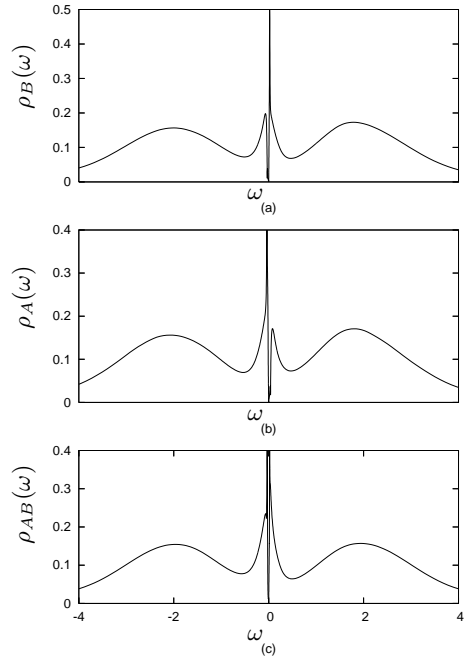}

 \caption{Three dimensional DOS from DMFT+NRG for
  $\epsilon_A=-2.07$, $\epsilon_B=-1.89$,
   $U_A=U_B=3.96$, $T\to 0$. The NRG treatment of
  the ionic Hubbard model shows a clear quasi-particle hybridization
  gap at the Fermi level.
  For the NRG we used $\Lambda=2.3$ as discretization parameter
  \cite{Wilson75}  and kept $N_{st}=700$ states for each NRG
  iteration. 
  \label{fig:ABNRG}
}

\end{figure}

\begin{figure} [htbp] 
  \centering \psfrag{pi/2}{$ \frac{\pi}{2a}$}
  \psfrag{-pi/2}{$-\frac{\pi}{2a}$} \psfrag{z}{$\omega$}
  \psfrag{rhoA}{$\rho_A(k,\omega)$} \psfrag{rhoB}{$\rho_B(k,\omega)$}
  \psfrag{rhoAB}{$\rho_{AB}(k,\omega)$} \psfrag{k}{$ k$}
  \includegraphics [scale=0.5]{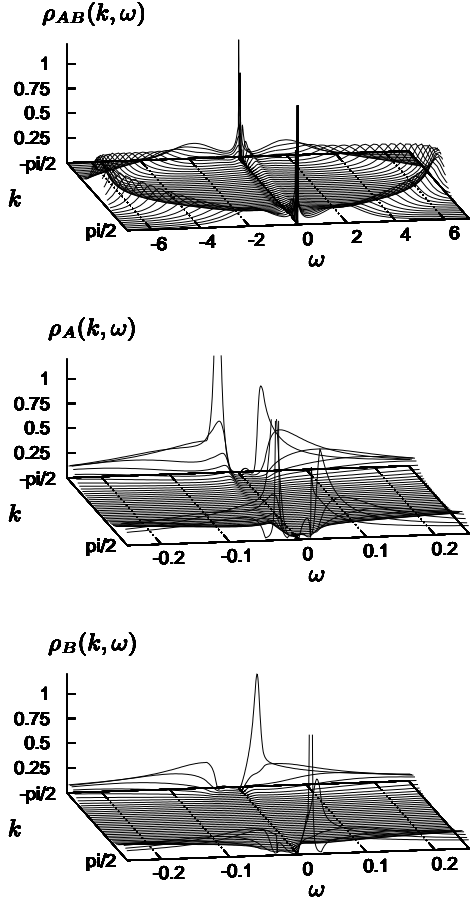}
\caption{[111] k resolved DMFT+NRG spectra with the same parameters as in
 Fig.~\ref{fig:ABNRG}. The top picture shows the whole spectrum for the
  combined DOS while the other two picures show the partial DOS for
  the A and B sublattice
  for the region around the quasiparticle bands at the fermi level. 
  \label{fig:knrgspek}
}

\end{figure}

\noindent
We have suggested above, that the quasi-particle states are$ \,$ shaped $ \,$by
hybridization $\,$ in $\,$  a$ \,$  similar $\, $ way $\,$ as  the\\
Hubbard-split bands of width $ W \propto t$
in the high energy region. It would therefore be consistent, if the size of this gap compared
to the one seen e.g. in Fig. 3 is scaled down roughly by the same factor of 
$\gamma =T^{\star}/W$ as the whole band structure. $T^{\star} $ is a characteristic
low energy scale caused by the strong correlations, which should be roughly of the same
magnitude as the enhanced Kondo scale of the nonionic version of the model.
As was pointed out in section 2 the gap
in Free Theory equals $\Delta$, so that a quasi-particle gap size $\Delta^{\star}$ of order
$\gamma \, \cdot \Delta$ can be expected. Whereas this order of magnitude actually is found in
our calculation, and
$\Delta^{\star}$ clearly vanishes when $\Delta$ goes to zero and increases with growing $\Delta$, 
a strict proportionality is not to be expected. Lifetime effects in the DMFT calculation start 
to smear features in the DOS when the distance to the Fermi level becomes larger and tend to narrow
a gap situated there. In addition, the existence of a quasi-particle gap causes a change in the scale
$T^{\star}$ as compared with the nonionic case.
 It would, however, be interesting to study the variation of
$\Delta^{\star}$ near the metal-insulator transition or away from half-filling, although
one has to expect numerical difficulties. A further similarity concerns the distribution of spectral
weight in the partial DOS, which formerly was discussed in connection with the structures
at higher energies. Even without $\k$-resolution the spectral weight for the A-lattice 
near $\omega=0$ in
Figs.\hspace{1mm}5(a) and \hspace{1mm}5(b) resembles the asymmetric
shape observed near the hybridization gaps at higher excitation energy
in Figs.\hspace{1mm}4(a) and 4(b), and also the $\k$-dependence reveals strong similarities.
 This is demonstrated in
Fig.\hspace{1mm}6, where the upper part gives an overall view of the
total weight $\rho_{AB} (\k, \omega)$ and the two lower parts magnify
the region very near to the Fermi level $ \omega = 0$. \newline\indent
Regarding the regions at large excitation energies $\w$ in the upper
figure, considerable broadening and spreading of the resonances is
observed compared to those of the Free Theory, see e.g.
Fig.\hspace{1mm}3. Moreover, band splitting due to the ionic field
$\Delta = \e_B - \e_A$ is not visible anymore and it seems that parts
of the broadened band structure even have disappeared. Although the
NRG treatment looses much accuracy away from the Fermi level due to
the logarithmic discretization of energies, we attribute these
findings to the dominance of scattering processes in this region due
to the blocking effect on the effective sites.
%Zitat
For the ionic version of the Hubbard model this seems to be of
particular importance.  Without the ionic field $\Delta$ one may
define the positions $\w_{\k}$ of broadened bands as usual
\cite{Luttinger1960} via the disappearance of the real part in the
denominator of one particle Greens functions, i.e. solving for $ \Re e
\; \tilde{G} (\w_{\k} + i \delta)^{-1} - \e_{\k} \equiv \w_{\k} -
\left( \e_{\k} + \frac{U}{2} + \Re e \; \tilde{\Sigma}(\w_{\k})\right)
= 0 $ in the Hubbard model, where we have separated a Hartree-part
$\frac{U}{2}$ from the selfenergy of the effective Greens function
$\tilde{G}(z)$ and explicitly used its $\k$-independence in the
effective site picture. Although $\Re e \; \tilde{\Sigma} (\w + i
\delta)$ bears a strong $\w$-dependence, generally three solutions
appear for the Hubbard model in the Fermi liquid regime which
represent the two original $U$-split bands and the low-lying
quasi-particle band. Level broadening occurs separately via $Im
\tilde{\Sigma} (\w_{\k} + i \delta)$. For the ionic Hubbard model the
denominator of the Greens functions \cite{BrandtMielsch89} mixes
contributions of the two kinds of effective sites in a way that
stresses the importance of scattering even more, i.e. one has to solve
\begin{eqnarray}
&&\left(\w_{\k} -\varepsilon_A - \Re e  \; \tilde{\Sigma}_A(\w_{\k})\right)
\left(\w_{\k} -\varepsilon_B - \frac{U}{2} - \Re e \tilde{\Sigma}_B (\w_{\k})\right) \nonumber \\
&& - Im \tilde{\Sigma}_A (\w_{\k} + i \delta) \cdot Im \tilde{\Sigma}_B (\w_{\k}  + i \delta)
- \e_{\k}^2 = 0.
\end{eqnarray}
Roughly speaking, $Im \tilde{\Sigma}_A (\w_{\k} + i \delta) \approx Im
\tilde{\Sigma}_B (\w_{\k} + i \delta) \approx \Delta \e $ is a large
energy shift for $\w_{\k} $ of order $ \e_A, \e_A + U$ , so that only
the solutions, which are maximally removed from the Fermi level tend
to remain in the high energy region after inclusion of the imaginary
part of the self energies. In the low energy region, on the other
hand, this effect is smallest near the Fermi level, which around
half-filling makes the zone boundaries favourable for the existence of
well defined quasi-particle bands. Thus we observe more pronounced
resonances there as shown in the two lower parts of Fig.\hspace{1mm}6.

Apparently, the dominant parts of the quasi-particle bands reproduce
the distribution of main spectral weight on the respective sublattice.
The parts on the opposite side of the Fermi level, on the other hand,
are shaped by admixtures, e.g. the spectral weight at the zone
boundaries does not vanish anymore. Near the position of the ionic
energies the NRG treatment combined with the effects described before
leads to a less detailed band structure.  This region is presumably
described in a somewhat better way by the NCA calculation, see e.g.
Fig. 7.  Thus, our expectations that fundamental hybridization
mechanisms, contained in the model, bear a relevance on the low lying
states with many-body character, is supported. We have compiled SNCA
and NRG results in our last Fig.\hspace{1mm}7 in order to provide an
overall picture of the local DOS and both of its sublattice
contributions. This also serves to underline our remarks about the
usefulness of both calculational schemes in conjunction.

\begin{figure} [ht] 
  \centering
  
  \psfrag{energie}{$\scriptstyle \omega$} \psfrag{rhoA}{$\scriptstyle
    \rho_A(\omega)$} \psfrag{rhoB}{$\scriptstyle \rho_B(\omega)$}
  \psfrag{rhoAB}{$\scriptstyle \rho_{AB}(\omega)$}
  \psfrag{xxxNCA}{$\scriptstyle SNCA$} \psfrag{xxxNRG}{$\scriptstyle
    NRG$}
  
  \includegraphics
  [scale=0.45]{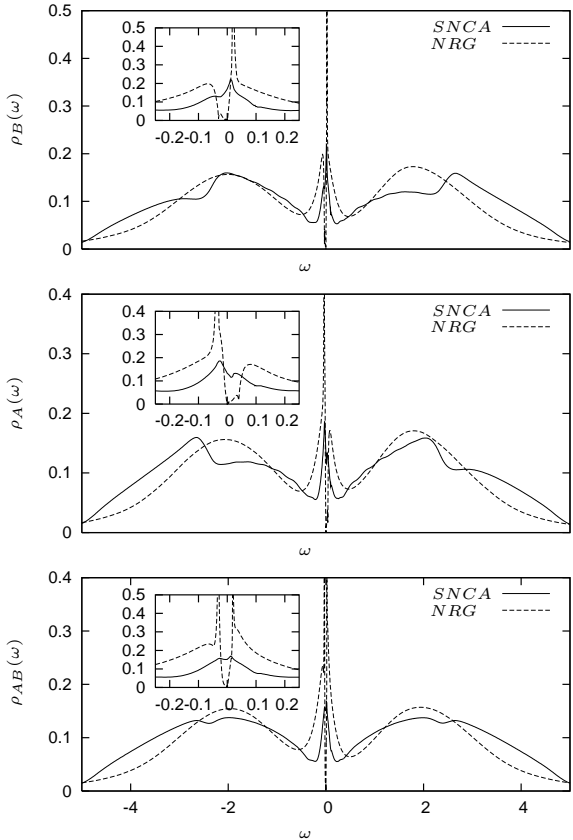}
 \label{fig:ABNCA}
 
 \caption{Comparisons  DOS obtained from DMFT with
   NRG and NCA for a $3d$ simple cubic lattice. The insets  show
   the region around the Fermi  level in greater detail. The
   parameters used are the same as in 
   Fig.~\ref{fig:ABNRG}; for the NCA a finite
   temperature $1/T=\beta=20$ was used. The NCA resolves much better the high energy
   features of the spectra but is not able to clearly resolve the
   quasiparticle hybridization gap as  the NRG. }
\end{figure}

\maketitle\section{Conclusion} We have calculated correlated band
structures of the ionic Hubbard model for a region of intermediate
local repulsion at and near half-filling. The quasi-particle bands of
low lying one-particle excitations show clear signatures of the
underlying transfer ("hybridization") mechanism contained in the
noninteracting part of the Hamiltonian, in particular with respect to
a charge excitation gap and the distribution of spectral weight. We
have motivated and interpreted these findings with reference to the
Luttinger sum rule and to the dominantly local nature of the
selfenergy and have drawn some parallels to gap formation in
the Anderson lattice. Applying successively better approximations we tried to
elucidate the transformation from a local ionic picture to essentially
itinerant quasi-particle degrees of freedom and the correspondance to
an intermediate tight binding scenario.

Using the SNCA
in the DMFT selfconsisting cycle gave a good overall impression of the
band structure and allowed for some qualitative conclusions about the
low lying excitations.  A satisfactory picture of the low energy
region at low temperatures was nevertheless only achieved by
application of the numerical renormalization group. Whether a full or
improved NCA would allow for a more analytical approach and would lead
to sensible results in this region, too, is still to be demonstrated.
Altogether, the scope of controlling and understanding such
calculations for complicated systems with more ionic species and
realistic p- or d-shells, in the local approach or e.g. in the
LDA-DMFT scheme \cite{HeldNekrasovBluemerAnisimovVollhardt2001}, has
been widened and improved: Some features of heavy quasi-particles seem
definitely closely connected to a noninteracting system. Since the
whole field at present rapidly develops to a stage, which is
technically very much involved, this insight should be useful for
future calculations. \newline\indent As far as the existence of
additional long ranged correlations and corresponding order parameters
are concerned, a possible instability of the Fermi-liquid state
considered here towards magnetic order or towards phase separation
should be investigated. Like in the homogeneous Hubbard model
\cite{JarrellPruschke1993} knowledge of the one particle excitations
is an important prerequisite for this program. Work on two particle
properties within the DMFT approach, and in particular on magnetic
and charge susceptibilities \cite{SchmittGrewe}, is in progress. Without further
calculations our results also demonstrate an asymmetry of occupation
numbers $n_A - n_B \neq 0$ away from half-filling
\cite{PozgajcicGross2003}, as a consequence of the asymmetric
distribution of spectral densities, see e.g. Fig.\hspace{1mm}4(a, b).


\begin{thebibliography}{99}
\bibitem{Hubbard81} J. Hubbard and J. B. Torrance: Phys.~Rev.~Lett.
  {\bf 47}, 1750 (1981)
\bibitem{PozgajcicGross2003} K. Pozgajcic and C. Gros: Phys.~Rev.~B
  {\bf 68}, 085106 (2003)
\bibitem{WilkensMartin2001} T.~Wilkens and R.~Martin: Phys.~Rev.~B
  {\bf 63}, 235108 (2001)
\bibitem{MedenSchoenhammer2003}S.~R.~Manmana, V.~Meden, R.~M.~Noack
  and K. Sch\"onhammer: arXiv: cond-mat/0307741v1 (2003)
\bibitem{ZitzlerPruschkeBulla2002} R. Zitzler, Th. Pruschke and R.
  Bulla: Euro. Phys. J. B {\bf 27}, 473 (2003)
\bibitem{GreweSteglich1991} N. Grewe and F. Steglich: Handbook Phys.
  and Chem. Rare Earths, Eds. K. A. Gschneidner jr. and L. Eyring,
  Elsevier, Amsterdam {\bf 14}, 343 (1991)
\bibitem{StichtD'AmbrumenilKuebler1986} J. Sticht, N. D'Ambrumenil and
  J. K\"ubler: Z. Phys. B {\bf 65}, 149 (1986)
  
\bibitem{PruschkeJarrellFreericks1995} Th. Pruschke, M.  Jarrell and
  J.  K.  Freericks, Adv. in Phys {\bf 42}, 187 (1995)
  
\bibitem{GeorgesKotliarKrauthRozenberg1996} A. Georges, G. Kotliar, W.
  Krauth and M. J. Rozenberg: Rev. Mod. Phys. {\bf 68}, 13 (1996)
  
\bibitem{BrandtMielsch89} U. Brandt and Chr. Mielsch, Z. Phys. B {\bf
    75}, 365 (1989)
  
\bibitem{HeldNekrasovBluemerAnisimovVollhardt2001} K. Held, I. A.
  Nekrasov, N. Bl\"umer,
  V. I. Anisimov and D. Vollhardt: Int. J. Mod. Phys. B {\bf 15}, 2611 (2001);\\
  M. B. Z\"olfl, Th. Pruschke, J. Keller, A. I. Poteryaev, I. A.
  Nekrasov and V. I. Anisimov: Phys. Rev. B {\bf 61}, 12810 (2000)
\bibitem{Luttinger1960} J. M. Luttinger: Phys. Rev. {\bf 119}, 1153
  (1960)
\bibitem{MarinAllen1979} R. M. Martin and J. W. Allen: J. Appl. Phys. {\bf 50}, 7561 (1979);\\
  R. M. Martin: Phys. Rev. Lett. {\bf 48}, 362 (1982)
\bibitem{Grewe1987} N. Grewe: Z. Physik B - Cond. Matter {\bf 67}, 323
  (1987)
\bibitem{Grewe1984} N. Grewe: Solid State Commun. {\bf 50}, 19 (1984)
\bibitem{FuldeKellerZwicknagl1988} P. Fulde, J. Keller and G.
  Zwicknagl: Solid State Phys., Eds.  H. Ehrenreich and D. Turnbull,
  Academic Press, San Diego {\bf 41}, 1 (1988)
\bibitem{Kuramoto1985} Y.Kuramoto: Theory fo Heavy Fermions and
  Valence Fluctuations,
  Eds. T. Kasuya and T. Saso, Springer, Berlin, 152 (1985); \\
  Ch. Kim, Y. Kuramoto and T. Kasuya: J. Phys. Soc. Japan {\bf 59},
  2414 (1990)
\bibitem{MetznerVollhardt1989} W. Metzner and D. Vollhardt: Phys. Rev. Lett. {\bf 62}, 324 (1989);\\
  W. Metzner: Physica B {\bf 165} \& {\bf 166}, 403 (1990);\\
  D. Vollhardt: Correlated Electron Systems, Ed. V. Emery, World
  Scientific, Singapore (1993)
\bibitem{PruschkeBullaJarrel2002} Th. Pruschke, R. Bulla and M. Jarrel: Phys. Rev. B {\bf 61}, 12799 (2000);\\
  R. Bulla, T. A. Costi and D. Vollhardt: Phys. Rev. B {\bf 64},
  045103-1 (2001)
\bibitem{PruschkeGrewe1989} Th. Pruschke and N. Grewe: Z. Physik B -
  Cond. Matter {\bf 74}, 439 (1989)
\bibitem{KeiterKimball1971} H. Keiter and J. C. Kimball: Int. J. Magn. {\bf 1}, 233 (1971);\\
  N. Grewe and H. Keiter: Phys. Rev. B {\bf 24}, 4420 (1981)
\bibitem{Metzner1991} W. Metzner: Phys. Rev. B {\bf 43}, 8549 (1991)
\bibitem{Grewe1983} N. Grewe: Z. Physik B - Cond. Matter {\bf 53}, 271
  (1983)
\bibitem{KuramotoKojima1984} Y. Kuramoto and H. Kojima: Z. Physik B -
  Cond. Matter
  {\bf 57}, 95 (1984);\\
  E. M\"uller-Hartmann: Z. Physik B - Cond. Matter {\bf 57}, 281 (1984);\\
  Y. Kuramoto and E. M\"uller-Hartmann: J. Magn. Mat. {\bf 52}, 122
  (1985)
\bibitem{AndersGrewe1994} F. B. Anders and N. Grewe: Europhys. Lett.
  {\bf 26}, 551 (1994)
\bibitem{KrohaWoelfleCosti1997} J. Kroha, P. W\"olfle and T. A. Costi:
  Phys. Rev. Lett. {\bf 79}, 261 (1997)
  
\bibitem{Wilson75} K.~G. Wilson, Rev. Mod. Phys. {\bf 47}, 773 (1975).
  
\bibitem{KrishWilWilson80b} H.~R. Krishna-Murty, J.~W. Wilkins, and
  K.~G. Wilson, Phys. Rev. B {\bf 21}, 1044 (1980).

%Zitate zur energieabhängigen DOS in der NRG
\bibitem{BullaPruschkeHewson97} R. Bulla, Th. Pruschke and A. C.
  Hewson, J. Phys.: Condens. Matter {\bf 9}, 10463 (1997).

%Zitat zur Berechnung der lokalen Greenschen Funktion
\bibitem{BullaHewsonPruschke98} R. Bulla, A.~C. Hewson, and T.
  Pruschke, J. Phys.: Condens. Matter {\bf 10}, 8365 (1998).
  
\bibitem{JarrellPruschke1993} M. Jarrell and Th. Pruschke: Z. Physik B - Cond. Matter {\bf 90}, 187 (1993);\\
  Th. Pruschke and R. Zitzler: arXiv: cond-mat/0309192v1 (2003)
\bibitem{SchmittGrewe} S. Schmitt and N. Grewe: to be publ. in Proc. 
Intern. Conf. on Strongly Correlated Electron Systems 2004, Physica B

\bibitem{This term}This term also seems
appropriate for the following reason: The Hubbard I-approximation
may be reformulated with a perturbation expansion with respect
to hopping or hybridization to neighbours. In this frame the Free Theory
encorporates 
all processes not containing local cummulant vertices and realizes
as such a very general approximation scheme in which Wicks theorem
is formally applicable and propagation through the lattice is free.
\end{thebibliography}
\end{document}